\documentclass[preprint,prX]{revtex4}

\usepackage{epsfig,graphicx}
\usepackage{array}

\newcommand{\dps}{\displaystyle}
\begin{document}

\title{On alpha stable distribution of wind driven water surface wave slope}

\author{M. Joelson and M.C. N\'eel}
\affiliation{UMR EMMAH, INRA Domaine St Paul, 84914 Avignon cedex
9, France}

\email{maminirina.joelson@univ-avignon.fr}
\date{\today}
{}


\begin{abstract}
We propose a new formulation of the probability distribution
function of wind driven water surface slope with an
$\alpha$-stable distribution probability. The mathematical
formulation of the probability distribution function is given
under an integral formulation. Application to represent the
probability of time slope data from laboratory experiments is
carried out with satisfactory results. We compare also the
$\alpha$-stable model of the water surface slopes with the
Gram-Charlier development and the non-Gaussian model of Liu et
al\cite{Liu}. Discussions and conclusions are conducted on the
basis of the data fit results and the model analysis comparison.\\
\\
{\bf Keywords: }{\it Wind driven water surface wave slopes.
Probability density function. Non-Gaussian distribution.
Alpha-stable distribution.}\\
{\bf During the last three decades, remote sensing of ocean
surface has been extensively investigated to study physical
processes on the sea surface and air sea interaction mechanism.
One of the most important quantities on which depends the ocean
remote sensing process is the surface wave slope probability
distribution that is known to be non-Gausssian. In the past the
general way to describe such probability distribution was to use
an expansion technique of the Gaussian law known as Gram-Charlier
series. In the present paper, we propose a new formulation of the
surface wave slope using alpha stable probability distribution.
Application to data from laboratory experiments and comparison of
the alpha stable slope model with existing models are carried out.
Our results show the potentiality of the alpha stable laws to
describe the non-Gaussian variability of water surface wave
slope.}
\end{abstract}

\maketitle

\section{Introduction:}

A knowledge of the sea surface slope distribution and its related
statistical quantities is essential to study the sea surface using
remote sensing processes. The use of radar backscatter to
characterize the sea surface has been widely investigated
experimentally and theoretically. At nearly normal radar beam
incidence, the importance of the slope distribution is clearly
exhibited by the simplest model of cross section radar backscatter
based on geometrical optics approach formulated by
Valenzuela\cite{Valenz} as:
\begin{equation}
\sigma_0(\theta)=\pi sec^4\theta f(\zeta_x,\zeta_y)|R(0)|^2
\label{valenzuela}
\end{equation}
where $\theta$ is the radar incidence angle, $|R(0)|^2$ is the
Fresnel reflection coefficient for normal incidence and
$f(\zeta_x,\zeta_y)$ is the probability distribution function
(PDF) of the sea surface slope at specular point
$(\zeta_x,\zeta_y)$. Determination of the surface slope
distribution has been carried out in a large part with
semi-empirical models derived from data observations. The commonly
used approach for such purpose was the truncated Gram-Charlier
series under the assumptions of the near-Gaussian case and the
weakly nonlinear interaction proposed by
Longuet-Higgins\cite{Longuet}. The Gram-Charlier series was
attempted to include two additional factors: the skewness and the
peakedness that are respectively due to second and third order
wave-wave interactions. Despite the multiple adaptations made on
the Gram-Charlier model, it is now well-known that this approach
fails to work in the range of large slope or large surface
elevation. In other word, this model is not suitable to represent
the reality when large events have higher occurrence. To improve
the results from Gram-Charlier series, several works were made by
different authors. Among the existing models, those of Liu et
al\cite{Liu}, differs by its formulation and its mathematical
nature. Indeed, Liu model does not appeal to an expansion
technique and is well suitable to represent extreme events.
However, this model is originally a symmetrical law including the
Gaussian probability and the Cauchy law as limit cases. Owing to
the symmetry property, the Liu model is not able to reproduce the
skewness effect. Liu extended empirically the model for this
purpose.

In the present paper we propose to use the $\alpha$-stable law to
describe the PDF of surface slope. The stable law depends on four
parameters: the index of stability, the skewness, the scale and
the location parameters that allow one to represent a probability
law of random variations including asymmetry, peakedness and peak
shift. Also, the $\alpha$-stable law includes as limit cases the
Gaussian probability and the Cauchy law. Generally speaking,
stable law belongs to the family of heavy tails probability laws
that is especially well adapted to take in account the large
events of random variations. Such strong random variations can be
found in fields or laboratory observations of the water surface
roughness (Onorato et al\cite{Ono1}).

Evidences from in situ and laboratory experiments of non-Gaussian
heavy tails distribution of wind driven water surface waves and
their slope may be found in many articles in the literature. The
first argument on the non-Gaussian behavior of wind driven water
surface waves and their slopes concerns the existence of asymmetry
property of wind wavefield that can be quantified by higher order
spectral analysis (Leykin et al\cite{Leykin},Joelson et
al\cite{Joelson}). Another hallmark of this non-Gaussian behavior
is based on the power law observed in the spectrum of wind driven
water surface waves. We cite here some works emphasizing this
fact. Experiment observations was conducted by Huang\cite{Huang}
on time series of sea surface that suggest a random dynamic as a
spectral power law associated to fractal description. Mellen and
al\cite{Mellen} performed the study of random nonlinear dynamics
of laboratory wind waves by the L{\'e}vy index and the
co-dimension parameter that quantify the multifractal property of
the water surface. As pointed out by Solow \cite{Solow}, power law
behaviour can arise from random distributions having heavy upper
tail. A fundamental consequence is that random rough surfaces with
slowly decaying power spectral density can have infinite slope
variance (Warnick et al\cite{Warnick}). Such case of slope
behavior is clearly beyond the Gaussianity assumption and is only
reached by the family of heavy tail probability distribution
model. The $\alpha$-stable law belongs to this class of model.
Heavy tail distributions of wind driven water surface slope
obtained from experiment observations can be found in different
works (see for example Cox and Munk \cite{Cox1}, Lui et
al\cite{Liu}, Chapron et al\cite{Chapron}).

Direct simulation methods such as a Monte-Carlo simulation based
on nonlinear hydrodynamical model are able to reproduce
pseudo-random samples that present similar aspects. Recently,
Soriano et al\cite{Soriano} studied the effect of this
non-Gaussian character on the radar Doppler spectrum of the random
surface at grazing angle and concluded on the importance of the
non-Gaussian character. Despite these results and the fact that
stable laws have been discovered several decades ago, there are
only few applications concerning the $\alpha$-stable distribution
of sea surface. In particular, we cite here the work of
Gu\'erin\cite{guerin} on the scattering on random rough surface
with $\alpha$-stable law in which he gave an extension of the
expression (\ref{valenzuela}) for the $\alpha$-stable surface
roughness.

The present study, in our opinion, is the first attempt to apply
the stable probability law on random water surface slope
representation. The paper is organized as follow: in \S 2, we
recall the main properties of the $\alpha$-stable law. In \S 3,
the application of stable law to fit instantaneous slope of wind
driven water surface is presented. The \S 4 is devoted to the
comparison of the stable PDF with the Gram-Charlier model and the
Liu model. In \S 5, we present our conclusions on the results of
the study.

\section{Stable probability distribution}
The stable laws were introduced by Paul L\'evy \cite{Lévy} during
his investigations of the behavior of the sums of independent
identically distributed random variables. Then the theory on
stable laws was developed by many authors and laid out in many
books and articles. We recall here some basic results without
proof. A stable distribution law is determined by four parameters:
an index of stability $\alpha$, a location parameter $\mu$, a
skewness parameter $\beta$ and a scale parameter $\gamma$. The
most common formulation of such law is given by the characteristic
function which represents the inverse Fourier transform of the
probability distribution function.
 A random variable X is said to have a stable distribution (noted as
 $\mbox{X} \sim S_{\alpha}(\beta,\gamma,\mu)$) if its characteristic
 function satisfies
\begin{equation}
\Phi_{X}(t)=\mbox{exp}\left(i\mu t -\gamma^{\alpha} |t|^{\alpha}
\left[ 1-i\beta \mbox{ sign}(t) \mbox{ W}(\alpha,t)\right] \right)
\label{f_c}
\end{equation}
where
$$
\mbox{W}(\alpha,t)= \left\{
\begin{array}{rl}
\mbox{ tan}(\frac{\pi\alpha}{2})  & \mbox{if } \alpha \neq 1
\\
-\frac{2}{\pi} \mbox{log } |t| & \mbox{if  } \alpha = 1
\end{array}
\right.
$$
and
$$
\mbox{sign}(t)= \left\{
\begin{array}{rl}
\mbox{1}  & \mbox{if } t > 0
\\
\mbox{0}  & \mbox{if } t = 0
\\
\mbox{-1}  & \mbox{if  } t < 0
\end{array}
\right.
$$
The stability index also called the characteristic exponent or the
tail index satisfies $0<\alpha\leq 2$ and the skewness parameter
$-1\leq\beta\leq 1$. The location parameter $\mu$ is defined only
when the stability index is $\alpha>1$ ($\mu \in \mbox{\bf{R}}$).
In this case, the first absolute moment is given by
$<|X|>=\frac{2\gamma}{\pi}\Gamma(1-1/\alpha)$ where $\Gamma(.)$ is
the gamma function. If $\beta=0$, then the distribution is
symmetric around $\mu$. The scale parameter also known as
dispersion parameter has a real positive value $\gamma>0$ that
determines the width of the distribution law. The location
parameter $\mu$ describes the shift of the peak. The tail index
$\alpha$ determines the rate at which the tails of the
distribution taper off. When $\alpha<2$, the variance of the
random variable X is infinite and the tails exhibit a power-law
behavior. More precisely, using the central limit theorem, it can
be shown that:
\begin{equation}
\left\{
\begin{array}{rl}
\lim\limits_{t \rightarrow \infty} t^\alpha \mbox{Pr}(X>\quad
t)=C_{\alpha}(1+\beta)\gamma^\alpha
\\
\lim\limits_{t \rightarrow \infty} t^\alpha
\mbox{Pr}(X<-t)=C_{\alpha}(1-\beta)\gamma^\alpha
\end{array}
\right. \label{tail}
\end{equation}
where
$C_{\alpha}=\frac{1}{\pi}\Gamma(\alpha)\mbox{sin}(\frac{\pi\alpha}{2})$.
It follows from (\ref{tail}) that in general, the
$\mbox{p}^{\mbox{th}}$ order moment
of a stable random variable is finite if and if only $p<\alpha$.\\

{\bf Probability density function}\\
There are only three cases of stable distributions for which a
closed-form expression of the PDF is known.
 This lack of closed form formulas for the PDF has a negative consequence on the popularity
 of the stable distribution. By applying Fourier inversion to
 (\ref{f_c}), an integral form of the PDF is obtained as:
\begin{equation}
\dps f_{\alpha,\beta,\gamma,\mu}(x)= \left\{
\begin{array}{rl}
\frac{1}{\pi}\quad\dps\int_0^\infty\mbox{exp}(-\gamma^\alpha
z^\alpha) \mbox{ cos} \left[(\mbox{x}-\mu)z+\beta z^\alpha\mbox{
tan}(\frac{\alpha\pi}{2})\right]dz
\\
\mbox{if } \alpha \neq 1
\\
\frac{1}{\pi}\quad\dps\int_0^\infty\mbox{exp}(-\gamma z) \mbox{
cos} \left[(\mbox{x}-\mu)z+\beta z \frac{\pi}{2}\mbox{
log}(|z|)\right]dz
\\
\mbox{if  } \alpha = 1
\end{array}
\right. \label{integ_form}
\end{equation}
Expression (\ref{integ_form}) is known to have analytical solution
for the following cases:\\
 -for $\alpha=2$ corresponding to the Gaussian distribution,
\begin{equation}
f_{2}(x)=\frac{1}{\sqrt{2\pi}\gamma}\mbox{exp}\left\{-\frac{(x-\mu)^2}{2\gamma^2}
\right\} \label{gauss}
\end{equation}
in this case, the skewness parameter is irrelevant. The variable
$\mbox{X} \sim S_{2}(0,\gamma,\mu)$ reduces to the Gaussian random
variable $\mbox{X} \sim {\cal N}(\mu,\sqrt{2}\gamma)$

 -for $\alpha=1$ and $\beta=0$ which leads to the Cauchy distribution
\begin{equation}
f_{1}(x)=\frac{1}{\pi} \frac{\gamma}{\gamma^2+ (x-\mu)^2}
\label{cauchy}
\end{equation}
The Cauchy random variable writes as $\mbox{X} \sim
S_{1}(0,\gamma,\mu)$

 -for $\alpha=0.5$ and $\beta=1$ giving the L\'evy distribution
\begin{equation}
f_{1/2}(x)=\sqrt{\frac{\gamma}{2\pi}}\quad\frac{\mbox{exp}\left\{-\frac{\gamma}{2x}\right\}}{x^{3/2}}
\end{equation}

For the general cases, numerical approximations of
(\ref{integ_form}) was widely used by several authors to represent
stable PDF. Different methods exist for this purpose with various
level of accuracy. In this paper, we adopt the direct integration
method developed by Nolan\cite{Nolan} and Weron\cite{Weron} using
an other formulation of the expression (\ref{integ_form}) called
the Zolotarev's formulas \cite{zol}. The main interest of the
Zolotarev's formulas is that instead of expression
(\ref{integ_form}), they do not include infinite integral and then
are well adapted to numerical computations. In which follow, we
give the Zolotarev's formulas of the PDF of a stable random
variable $\mbox{X} \sim S_{\alpha}(\beta,\gamma,\mu)$. First, we
consider the case of a random variable $\mbox{X}_0 \sim
S_{\alpha}(\beta,1,0)$. By setting $\zeta=-\beta \mbox{
tan}\frac{\pi\alpha}{2}$, the PDF of $\mbox{X}_0$ can be expressed
as:

$\mbox{if } \alpha \neq 1$
\begin{equation}
\dps f_{\alpha,\beta,1,0}(x)= \left\{
\begin{array}{rl}
\dps\frac{\alpha(x-\zeta)^{\frac{1}{\alpha-1}}}{\pi
|\alpha-1|}\dps\int_{-\zeta}^{\frac{\pi}{2}}\mbox{ V}
(\theta,\alpha,\beta)\mbox{
exp}\dps\left\{-(x-\zeta)^{\frac{\alpha}{\alpha-1}}\mbox{ V}
(\theta,\alpha,\beta)\right\} d\theta \mbox{,}
\\
\mbox{if } x> \zeta
\\
\dps\frac{\Gamma(1+\frac{1}{\alpha}) \mbox{ cos}(\xi)}{\pi}
\mbox{,} \qquad\qquad\qquad\qquad\qquad\qquad\qquad\qquad \mbox{if
} x = \zeta
\\
\dps f_{\alpha,-\beta,1,0}(-x) \mbox{,}
\qquad\qquad\qquad\qquad\qquad\qquad\qquad\qquad \mbox{ if } x <
\zeta
\end{array}
\right. \label{zolot}
\end{equation}

$\mbox{if } \alpha = 1$
$$
\dps f_{1,\beta,1,0}(x)= \left\{
\begin{array}{rl}
\frac{1}{2|\beta|} \mbox{ exp}( -\frac{\pi x}{2\beta})
\dps\int_{-\frac{\pi}{2}}^{\frac{\pi}{2}}\mbox{ V}
(\theta,1,\beta)\mbox{ exp}(-\mbox{ exp}( -\frac{\pi
x}{2\beta})\mbox{ V} (\theta,1,\beta)) d\theta \mbox{,} \quad
\\
\mbox{ if } \beta \neq 0 \quad
\\
\!\dps\frac{1}{\pi(1+x^2)}\mbox{,}
 \qquad\qquad\qquad\qquad\qquad\qquad\qquad\qquad
\qquad \mbox{if  } \beta = 0 \quad
\end{array}
\right.
$$
where $\xi=\frac{1}{\alpha}\mbox{ arctan}(-\zeta)$ and

$ \mbox{ V} (\theta,\alpha,\beta)= \left\{
\begin{array}{rl}
\left(\mbox{cos
}\alpha\xi\right)^{\frac{1}{\alpha-1}}\left(\frac{\dps\mbox{cos
}\theta}{\mbox{sin }\dps
\alpha(\xi+\theta)}\right)^{\frac{\alpha}{\alpha-1}}
\frac{\dps\mbox{cos }(\alpha\xi+(\alpha-1)\theta)}{\mbox{cos }\dps
\theta} \mbox{,     }\qquad\quad
\\
\alpha \neq 1\quad
\\
\dps\frac{2}{\pi}\left(\frac{\frac{\pi}{2}+\beta\theta}{\mbox{cos
}\dps \theta}\right)\mbox{exp}\left\{
\frac{1}{\beta}(\frac{\pi}{2}+\beta\theta)\mbox{ tan
}\theta\right\} \mbox{, }\quad\quad\quad\quad\quad \alpha = 1
\quad
\end{array}
\right. $

From the probability distribution given by (\ref{zolot}), one can
easily build the PDF of the random variable X with the help of the
stability property,$\mbox{ X}=\gamma\mbox{X}_0 +\mu \sim
S_{\alpha}(\beta,\gamma,\mu)$.

{\bf Parameters estimation}\\
The first step on modelling data with a stable law requires the
estimation of the four parameters that determine the distribution
properties. The estimation process suffers from the lack of known
closed form of the PDF. As a consequence, standard methods based
on maximum likelihood principle would not be efficient with
regards to the stable PDF dependance. However, there are other
numerical ways that do not depend directly on the PDF knowledge to
evaluate stable parameters like the sample quantile methods or the
sample characteristic function methods. A comparative analysis and
discussions about these methods may be found in Borak and
al\cite{borak}. Among these existing methods of estimation, we
will use the sample characteristic function developed by
Koutrovelis\cite{koutro} and performed by Kogon and
Williams\cite{Kog}. The method is based on regression type on the
log-characteristic function. Indeed, the logarithm of the real
part and the imaginary part of the log-characteristic function are
linear and then, give rise to a regression model with the data.
\begin{equation}
\begin{array}{rl}
\mbox{log }(-\mbox{Re}[\Psi(t)])=\alpha \mbox{ log} |t| + \alpha
\mbox{ log}\gamma
\\
\\
\mbox{Im}[\Psi(t)]=-\mu t -\beta \mbox{ sign}(t)\mbox{
W}(\alpha,t)
\end{array}
\end{equation}
where $\Psi$ is the log-characteristic function of X and W(.,.) is
defined as in expression (\ref{f_c}). A more complete description
of the method and an extended study may be found in \cite{Kog}.

In the following section, the stable PDF model will be applied on
slope data from laboratory experiments on wind driven surface
water waves.

\section{Measurements and data analysis}

  {\bf Experiments}

\begin{figure}[h]
\includegraphics[width=16cm,height=3.5cm]{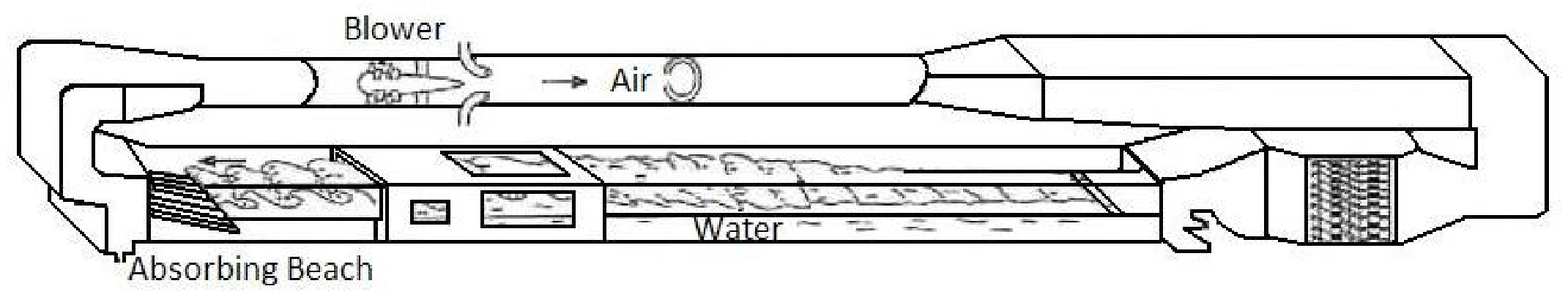}\vspace{0.2cm}
\caption{A schematic representation of the large air-sea
interactions facility.} \label{facility}
\end{figure}

The data of interest are issued from experiments made in the wind
wave facility of the IRPHE-IOA Laboratory. Basically, the facility
consists of a water tank of 40 m length, 3 m width and 1 m depth.
The facility can be described as a combination of a wind tunnel
with a wind-wave tank. A schematic representation of the air-sea
interaction simulation facility is given in Fig.\ref{facility}.
The facility is equipped with a submerged wavemaker that allows to
produce mechanical waves from capillary ripples to breaking waves.
However in the present work, we focus only on wind wave fields.
 By means of an axial fan, air flow with velocities ranging from
 0 m/s to about 15 m/s generates wind wave fields.
Different types of surface waves can be produced by the action of
the air flow. Noteworthy that the facility includes particular
devices intended to control various parasitical effects which
could coexist with the physical process under consideration. In
particular, a permeable wave absorber is set up at the upwind end
of the water tank to prevent the wave reflection. Within the range
of wave frequency of interest in the experiments on surface waves,
the reflection coefficient is estimated to be insignificant. As
far as wind generated waves are concerned, a specific design has
been adopted for a smooth joining of the air flow and the water
surface. The design prevents air flow-separation at the facility
entrance test section. This insures a natural development of the
turbulent boundary layer over the water surface. In addition,
later quays disposed in the water tank favor the three-dimensional
evolution of the wind waves. Studies with more details about these
specific devices are given in a number of publications (see e.g.
Coantic et al \cite{Coa}). We concentrate here on the time
evolution of the water surface deflection level at a given
position. Measurements of water surface height $\eta(\mbox{t})$
were performed by two capacitance wave gauges separated by a 5cm
along wind direction. These probes using a capacitance wire gauge
of 0.3 mm outer diameter were set up on a carriage moving along
the wave tank. Experiments were carried out under various
conditions, whose main parameters are the wind velocity and the
distance between the gauge and the facility entrance section
(hereafter denoted fetch). Such distance determines the length
action of the air flow over the water surface and is well known to
be of crucial importance on the rate growth of surface waves and
on its nonlinearity. The time derivative $\dps
\mbox{d}\eta\mbox{/dt}$ of the water surface heights were obtained
by using analog derivators.

The corresponding processed series are made of n = 36000 samples
gathered at a continuous rate of 200 Hz. The water surface
wavefields of interest here are known to have temporal as well as
spatial dynamics in which dispersion, nonlinearity, dissipation
(by kinematic viscosity or by breaking) and wind action are
involved. The results presented herein are limited to some aspects
of the temporal evolution at the given space locations. Thus,
instantaneous slope s(t) on the upwind direction of the water
surface waves could be derived as $\mbox{s(t)=-1/c}\dps \mbox{
d}\eta\mbox{/dt}$ where c corresponds to the phase velocity of the
dominant wave frequency. This slope measuring method was tested by
comparison with direct measurement performed by a laser slope
gauge (see Jhane and Riemer \cite{Jhane}). A good agreement was
found particularly for capillary and gravity waves at moderate
wind speed.

{\bf Data analysis}

In this section, we estimate the PDF of time water surface slopes
at different wind speed for fetch varying from 2m to 26m. Note
that wind speed is estimated at 10m height above the
surface(U$_{\mbox{\tiny{10}}}$). Among the large possibilities of
combinations from wind speed and fetch values, we present here
some cases that are indicative of the typical behavior of water
surface slope. In which follows, we focus on slope behavior of
three cases: capillary waves, capillary gravity waves at moderate
fetch and low wind speed and gravity waves at large fetch and high
wind speed that are typical of fully developed sea surface. In
order to compare the different runs of study, the data series are
normalized with theirs rms and mean values. The results of
the fit are summarized in Tab.\ref{tab_wave}.\\
\begin{table}
\begin{tabular}{lllllll}
\hline
Case &Fetch (m)& U $_{\mbox{\tiny{10}}}$ (m/s)  &
$\alpha$&$\beta$&$\gamma$&$\mu$\\
\hline
I&2&6&1.7219&-0.0816&0.5434&0.0056\\
II-1&6&6&1.8688 &-1.0000&0.6384& -0.0869\\
II-2&13&3&1.9822& -1.0000&0.6972 & -0.0117\\
III&26&13&1.9178 &-0.9488&0.6481&-0.0051\\
\hline
\end{tabular}
\caption{Estimation of stable parameters
($\alpha,\beta,\gamma,\mu$) from the data series.}
\label{tab_wave}
\end{table}

\begin{figure}[h]
\includegraphics[width=16cm,height=12cm]{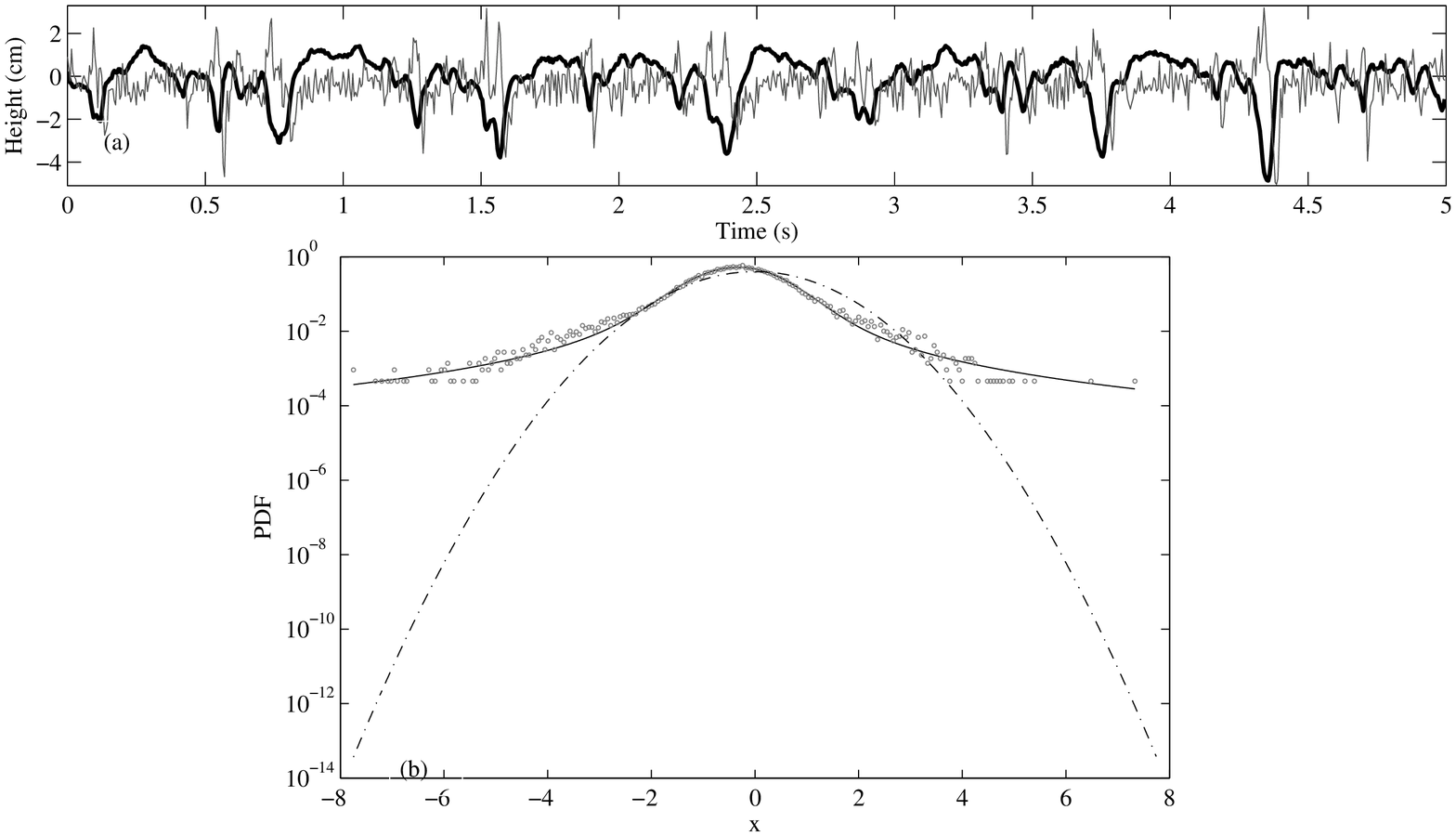}\vspace{0.2cm}
\caption{Fetch=2m. U$_{\mbox{\tiny{10}}}$ = 6m/s. Part(a): Thick
line: Normalized sample of surface wave. Thin line: Normalized
instantaneous slope sample. Part(b): 'o': PDF of slope experiment
data. Solid line: $\alpha$ stabe PDF. Dashed-dotted line:
Normalized Gaussian PDF} \label{signal_261}
\end{figure}

\begin{figure}[h]
\includegraphics[width=16cm,height=12cm]{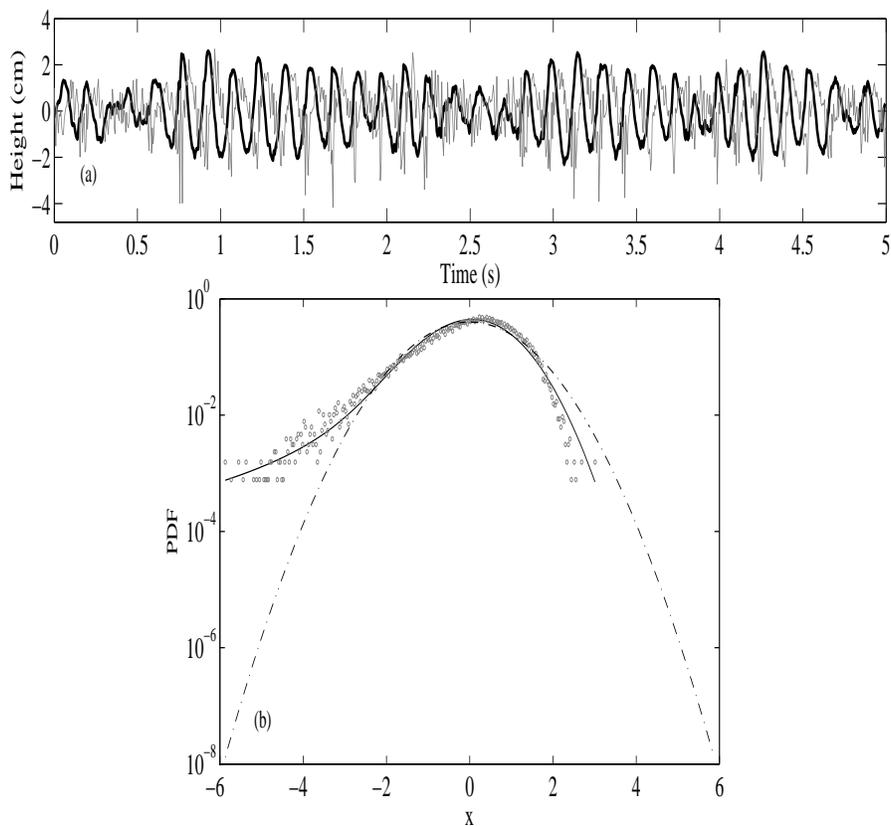}\vspace{0.2cm}
\caption{Fetch=6m. U$_{\mbox{\tiny{10}}}$ = 6m/s. Part(a): Thick
line: Normalized sample of surface wave. Thin line: Normalized
instantaneous slope sample. Part(b): 'o': PDF of slope experiment
data. Solid line: $\alpha$ stabe PDF. Dashed-dotted line:
Normalized Gaussian PDF} \label{signal_661}
\end{figure}

\begin{figure}[h]
\includegraphics[width=16cm,height=12cm]{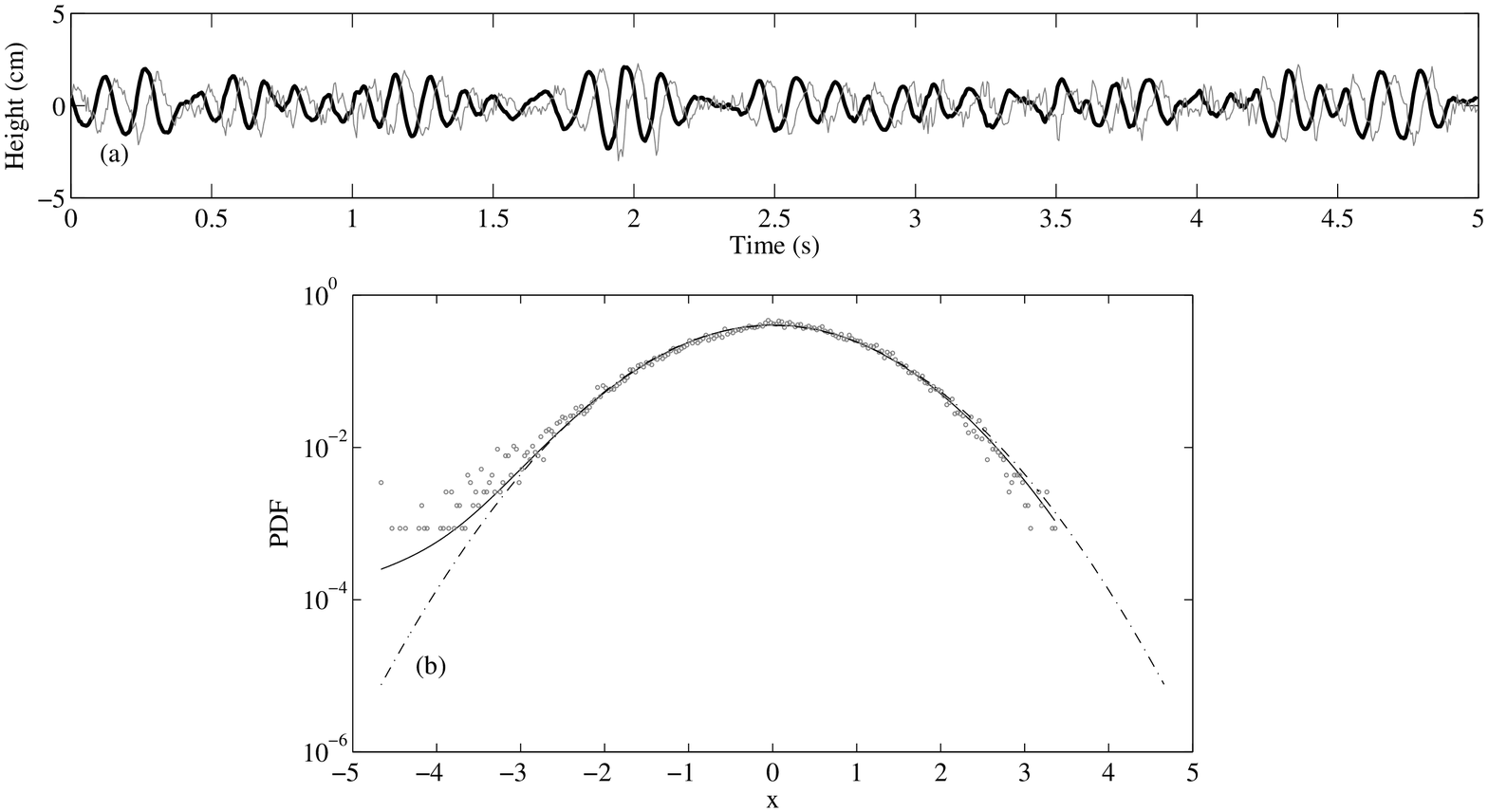}\vspace{0.2cm}
\caption{Fetch=13m. U$_{\mbox{\tiny{10}}}$ = 3m/s. Part(a): Thick
line: Normalized sample of surface wave. Thin line: Normalized
instantaneous slope sample. Part(b): 'o': PDF of slope experiment
data. Solid line: $\alpha$ stabe PDF. Dashed-dotted line:
Normalized Gaussian PDF} \label{signal_1331}
\end{figure}

\begin{figure}[h]
\includegraphics[width=16cm,height=12cm]{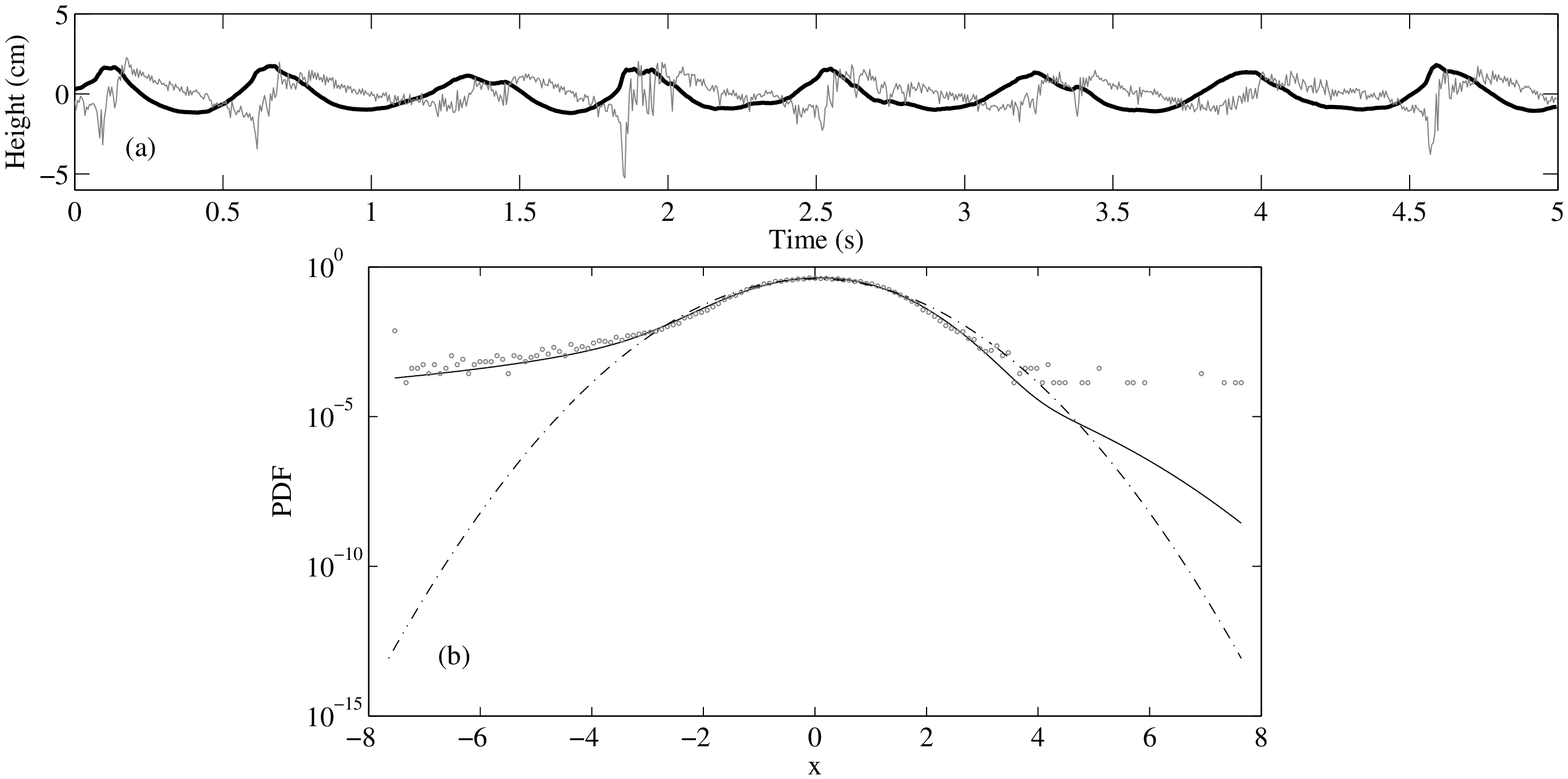}\vspace{0.2cm}
\caption{Fetch=26m. U$_{\mbox{\tiny{10}}}$ = 13m/s. Part(a): Thick
line: Normalized sample of surface wave. Thin line: Normalized
instantaneous slope sample. Part(b): 'o': PDF of slope experiment
data. Solid line: $\alpha$ stabe PDF. Dashed-dotted line:
Normalized Gaussian PDF} \label{signal_26131}
\end{figure}

The first line of Tab.\ref{tab_wave} corresponds to the case I of
capillary wind wave slopes at moderate wind speed and at a very
short fetch. In this case, we find a stability index of 1.72 that
is indicative of a strong non-Gaussian character. A very weak
value of the skewness parameter is also found traducing a lack of
asymmetry on the waveform. The scale parameter is at a relative
low value compared to the normalized Gaussian standard deviation($
\sim S_2(0,1/\sqrt{2},0)$). This may be related to a nonlinear
effect involved on the generation of capillary waves by wind. The
characteristics of the capillary wave slopes are shown in
Fig.\ref{signal_261}. The Fig.\ref{signal_261}.a depicts a sample
of the capillary wave profile and its slope while the
corresponding probability distribution is shown in the part (b) of
the figure with a normalized Gaussian law as a reference.
At larger fetch value corresponding to case II, the stability
index grows. The scale parameter also increases. A striking
feature is that the skewness parameter reaches to its negative
limit value. This is in good agreement with time evolution of the
slope  presented in Fig3.a. Observing also that the corresponding
surface waves are dominated by a Stokes component, we suggest the
possible role of the Stokes effect in the asymmetric behavior of
the slope and in the limit value obtained for the skewness
parameter in the present case. However, in the general case,
relation between asymmetry in the surface and in the slope
variation is not easy to establish especially for a random
wavefield. The Stokes effect yields a waveform with steeper crests
increasing locally slopes and flatter troughs. Results of run at
fetch of 6m and wind speed of 6m/s (case II-1) are shown in
Fig.\ref{signal_661}. The asymmetry property is clearly visible on
the slope samples (Fig.\ref{signal_661}.a) and is well represented
by the PDF model (Fig.\ref{signal_661}.b). Modulational
instabilities begin to occur in the wavefield owing to the
nonlinear wave-wave interaction processes as the Benjamin-Feir
instabilities. The sample of the wave signal presents wave trains
formation that are reminiscent of these instabilities effects (see
Fig.\ref{signal_661}.a). Such nonlinear effects are known in this
case to dominate the effects of randomness. This would explain the
reduction of slope magnitude and also the increase of the
stability index to the near-Gaussian situation. By increasing the
fetch value but taking a low wind speed (case II-2), the rate of
input wind energy will be weaker than the rate of the
self-nonlinearity of the wavefield. Thus, one can better put
forward the wave modulation effect on the slope distribution and
on its PDF. The corresponding PDF is shown in
Fig.\ref{signal_1331}.b with a very similar form to the normalized
Gaussian except for the negative tail that is indicative of the
presence of Stokes effect.
Finally, at high wind speed and large fetch corresponding to the
case III, one  can surprisingly find that values of the index
stability but also the skewness and the scale parameter reduce
reaching again a clearly non-Gaussian situation. In agreement with
these values, from the Fig.\ref{signal_26131}.a, it may be found
that the magnitude of the wave slopes grows locally at the crests
and the wave signal behaves strongly asymmetrical. From the
physical background, it was shown by Trulsen and
Dysthe\cite{Trulsen} that such situation may be explained from a
theoretical model based on a modified nonlinear Schr\"{o}dinger
equation. It allows one to predict the stabilization of the Stokes
effects by the strong wind leading to the suppression of the
modulational instabilities. The PDF of the case III are given in
Fig.\ref{signal_26131}.b in which it may be found that the stable
PDF model reproduce well the distribution of the negative slope
values and also the skewness of the slopes. However, at large
positive values, the model fails to represent the data
distribution. No clear interpretation was found for this
difference but we believe that additional physical processes such
as breaking wave would be involved.

 \clearpage

\section{Comparison with other models of slope distribution}
In this section, we compare the properties of the stable law with
the Gram Charlier and the Liu models. The Gram-Charlier model
based on an expansion technique called Edgworth series was widely
used in the past and permitted to develop a large part of the
understanding of the non-Gaussian behavior of the sea surface.
However, the limit of this model was earlier found by several
authors. Many efforts were conducted to develop alternative ways
from the statistical expansion technique. The Liu model was built
in this context to improve the statistical representation of the
sea surface slopes.

{\bf The Gram-Charlier series}

The traditional method of interpreting the wind wave slopes has
been to fit them to a Gram-Charlier series, which can be derived
by assuming that surface waves are weakly nonlinear
(Longuet-Higgins\cite{Longuet}). Basically, in this procedure the
PDF of the wave slopes s(t) writes as:
\begin{equation}
\phi_{\mbox{\tiny{GC}}}(s)=\mbox{G}(s)\left[1+\frac{\mbox{C}_3}{6}\mbox{
H}_3(s)+\frac{\mbox{C}_4}{24}\mbox{ H}_4(s)\right] \label{gram}
\end{equation}
where
$$
\mbox{G}(s)=\frac{1}{\sigma\sqrt{2\pi}}\mbox{ exp
}\left[-\left(\dps\frac{s-<s>}{\sqrt{2}\sigma}\right)^2\right]
$$
and $\sigma$ is the rms of the slope, when $\mbox{H}_n$ is the
$\mbox{n}^{\mbox{th}}$ Hermite polynomial. The coefficients
$\mbox{C}_n$ can be calculated from the statistical moments of
$\phi_{\mbox{\tiny{GC}}}$(s) and G(s) as:
$$
C_n=\frac{1}{\sigma^n}\dps\int s^n
\left[\phi_{\mbox{\tiny{GC}}}(s)-\mbox{G}(s)\right]ds
$$
Clearly, the coefficients $\mbox{C}_3$ and $\mbox{C}_4$ in
expression (\ref{gram}) are intended to include the skewness and
the kurtosis that are the consequences of nonlinearities in the
underlying physical processes. Of interest is that the PDF model
becomes non-Gaussian. Note also that expression (\ref{gram}) is
not always guaranteed to be positive, and is therefore not a valid
probability distribution. The Gram-Charlier series diverges in
many cases of interest. It converges only if
$\phi_{\mbox{\tiny{GC}}}$(s) falls off faster than
$\mbox{exp}(-s^2/4)$ at infinity (Cram\'er\cite{Cram}). For water
waves applications, the skewness and kurtosis coefficients ($C_3$
and $C_4$) are found to be related to physical parameters in
particular to the air flow speed value at 10m altitude (noted as
the $\mbox{U}_{10}$). Here we adopt the values of these
coefficients used by Cox \mbox{\&} Munk\cite{Cox} and
Plant\cite{Plant} from a range of wind speed varying from 0.72 to
10.2 m/s (Note that \cite{Cox} uses the wind speed at height of
12.5m instead of the traditional $\mbox{U}_{10}$). We fit the
Gram-Charlier model with stable law using the
Nelder-Mead\cite{Nelder} simplex method. The results of the fit
are given in Tab.\ref{tab1}. The method is based on the
minimization of the euclidian norm of the error $\epsilon=||\mbox{
f}_{\alpha,\beta,\gamma,\mu}-\phi_{\mbox{\tiny{GC}}}||$ for 1000
points varying from -5 to +5. From each case, values of the best
stable parameters and the error $\epsilon$ are shown.

\begin{table}[h]
\begin{tabular}{llllllllll}
\hline
U$_{\mbox{\tiny{12.5}}}$ (m/s) & $\sigma$ &
$\mbox{C}_3^{^{CM}}$& $\mbox{C}_4^{^{CM}}$&
$\alpha$&$\beta$&$\gamma$&$\mu$&$\epsilon$\\
\hline
0.72&0.0005&0.101&0.127&1.951&-0.330&0.695&-0.05& $1e^{-1}$\\
3.93&0.0098&0.003 &0.129 &1.957 &0.022&0.695& 0&$1.9e^{-2}$\\
4.92&0.0174&-0.080&-0.019&1.980&0.900&0.708&0.040&$8.6e^{-2}$\\
6.30&0.0170&-0.143&0.101 &1.974& 0.880&0.697 & 0.070 &$1.5e^{-1}$\\
8.00&0.0191&-0.156&0.173 &1.930 &0.260&0.690 &0.07 & $1.5e^{-1}$\\
10.2&0.0357&-0.283&0.128& 1.910& 0.400 &0.690 &0.13 &$2.7e^{-1}$\\
\hline
\end{tabular}
\caption{Estimation of stable parameters
($\alpha,\beta,\gamma,\mu$) from the Gram-Charlier series used by
Cox and Munk (1954) model} \label{tab1}
\end{table}
As a general remark for all cases, the values of the stability
index $\alpha$ are close to $2$ that traduces the near-Gaussian
assumption adopted in the Gram-Charlier model. The error
$\epsilon$ increases with the wind speed values. Rather than a
simple difference from the calculation, these error values reveal
a more profound departure on the nature of both models. More
precisely, at low wind speed, when the occurrence of large values
of slope would be rare, a better agreement is found in particular
around the peak of the PDF.
\begin{figure}[h]
\includegraphics[width=18cm,height=10cm]{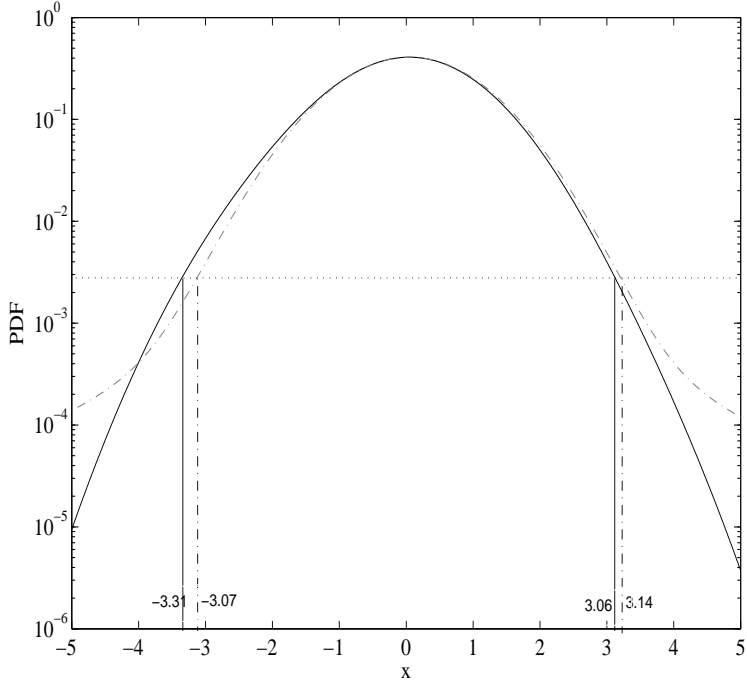}\vspace{0.2cm}
\caption{Results of the fit for U$_{\mbox{\tiny{12.5}}}$ =8m/s.
The solid line is the Gram-Charlier PDF. Dashed line represents
the stable law of the fit. A fixed value of the probability is
drawn with dotted line that allows one to visualize the asymmetry
of both laws.}  \label{cox_munk}
\end{figure}
At high wind speed, the results of the Gram-Charlier series differ
clearly from those of the stable law both at the tails of the
probability and at its middle part. The Gram-Charlier model
underestimates the extreme events giving a less heavy tails on the
probability law (see Fig.\ref{cox_munk}). At the middle part of
probability law, it overestimates the effect of the skewness. Note
that no clear tendency is found between the stable parameter
$\beta$ and the skewness $\mbox{C}_3^{^{\mbox{\tiny{CM}}}}$.
Finally, the location parameter $\mu$ is globally around zero and
the scale parameter $\gamma$ is found at a quasi constant value
around $\frac{1}{\sqrt{2}}$. They are directly related to the
normalized Gaussian law of the function G(s) included in the
expression (\ref{gram}).

 {\bf The Liu model of slope PDF}

A more advanced model is developed by Liu et al\cite{Liu} to fit
the joint probability of the surface wave slopes. Liu model is
also based on the weakly nonlinear assumption of the surface
\cite{Longuet}. The basic form of Liu slope PDF writes as:

\begin{equation}
\dps f_{L}^{n}(x,y)=\frac{n}{2\pi(n-1)}\left[ 1+
\frac{x^2}{(n-1)}+ \frac{y^2}{(n-1)}\right]^{-(n+2)/2}
\label{liu1}
\end{equation}
where x,y are normalized slope components and n an integer
parameter describing the peakedness of the surface slopes. Note
that (\ref{liu1}) refers to the case of isotropic sea surface and
is not always suitable for anisotropic real surface. An empirical
formulation inspired by \cite{Cox}\cite{Longuet} is used by
\cite{Liu} to handle the skewness effect. According to \cite{Liu},
the expression (\ref{liu1}) is an improvement of the Gram-Charlier
distribution in particular for the range of large slopes. For
convenience, here we consider the expression for y=0, i.e. we
focus only on upwind direction. As a first remark, the positivity
of the expression (\ref{liu1}) is obviously guaranteed however it
may be found that the expression of $f_{L}^{n}(x,0)$ does not
represent a valid probability density function for any value of n.
Indeed, the integral of this expression is not equal to unity. We
addres this point by adopting a normalized form of the Liu PDF as:
\begin{equation}
\phi_L^m(x)=
\frac{1}{\sqrt{\pi}}\left[1+\frac{x^2}{(m-1)}\right]^{(-\frac{m}{2}+1)}
\dps\frac{\Gamma(\frac{m}{2}-1)\sqrt{\frac{1}{m-1}}}{\Gamma(\frac{m}{2}-\frac{3}{2})}
\label{liu_mod}
\end{equation}
The expression (\ref{liu_mod}) is defined for $m\geq4$. For m=4,
one can easily find that
\begin{equation}
\phi_L^4(x)= \frac{1}{\pi}\left( \frac{\sqrt{3}}{3+x^2}\right)
\label{cauchy_1}
\end{equation}
Expression (\ref{cauchy_1}) belongs to the Cauchy stable law with
$\mbox{X} \sim S_{1}(0,\sqrt{3},0)$. On the other hand, taking the
logarithm of (\ref{liu_mod}), the limit for large value of m
writes as:
\begin{equation}
\mbox{log }\phi_L^m (x)\dps\simeq_{m\rightarrow \infty}
-\frac{m-2}{2}\mbox{ log}\left( 1+\frac{x^2}{m-1} \right)
\end{equation}
which reaches by simple series expansion to the normal
distribution of $\mbox{X} \sim {\cal N}(0,1)$ or also the stable
law with $\mbox{X} \sim S_{2}(0,\frac{1}{\sqrt{2}},0)$. The limit
cases of the Liu formulation belong to the symmetrical stable
distribution.

\begin{figure}[h]
\includegraphics[width=15cm,height=12cm]{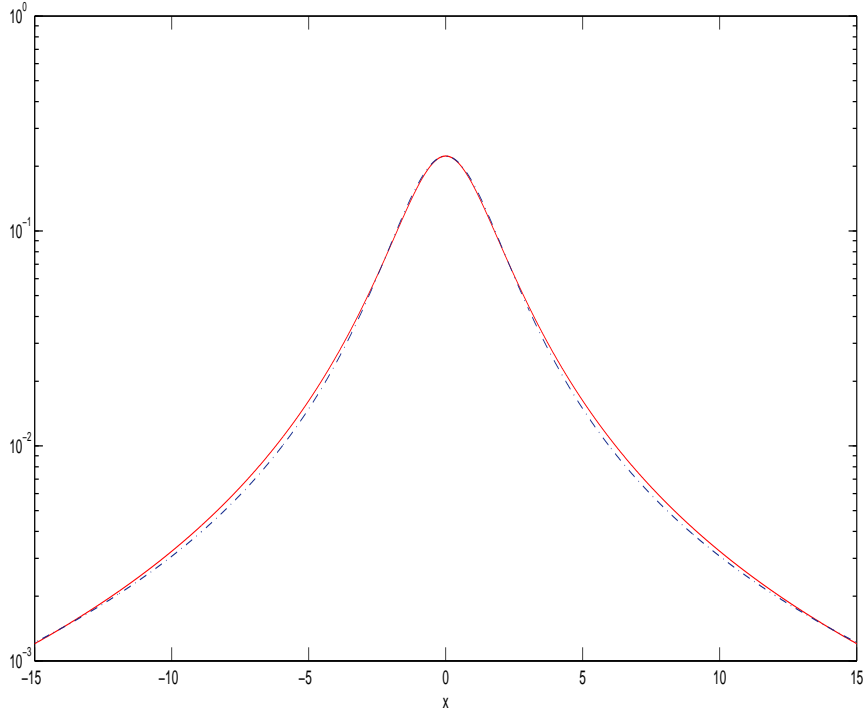}\vspace{0.2cm}
\caption{Semi-log representation of Liu PDF for m=4.5 and the
stable law fit with $\alpha=1.221$ and $\gamma=1.3384$. The error
$\epsilon=0.066277$. Solid line: Liu PDF model. Dashed line:
Stable law.} \label{liucompare}
\end{figure}

For intermediate values, the Liu PDF differs slightly from the
stable distribution. In order to illustrate this relation, we
compute a numerical adjustment of the stable parameters $\alpha$
and $\gamma$ to the expression (\ref{liu_mod}) for m=4.5. The
adjustment is based on the minimization of the euclidian norm
$||\mbox{ f}_{\alpha,0,\gamma,0}(x)-\phi_L^m(x) || \leq \epsilon$
for 3000 points from x=-15 to 15. The results are shown in
Fig.\ref{liucompare}. It may be found that the major difference
between the two laws are clearly localized at the medium range
values of the probability structure.

\begin{figure}[h]
\includegraphics[width=15cm,height=10cm]{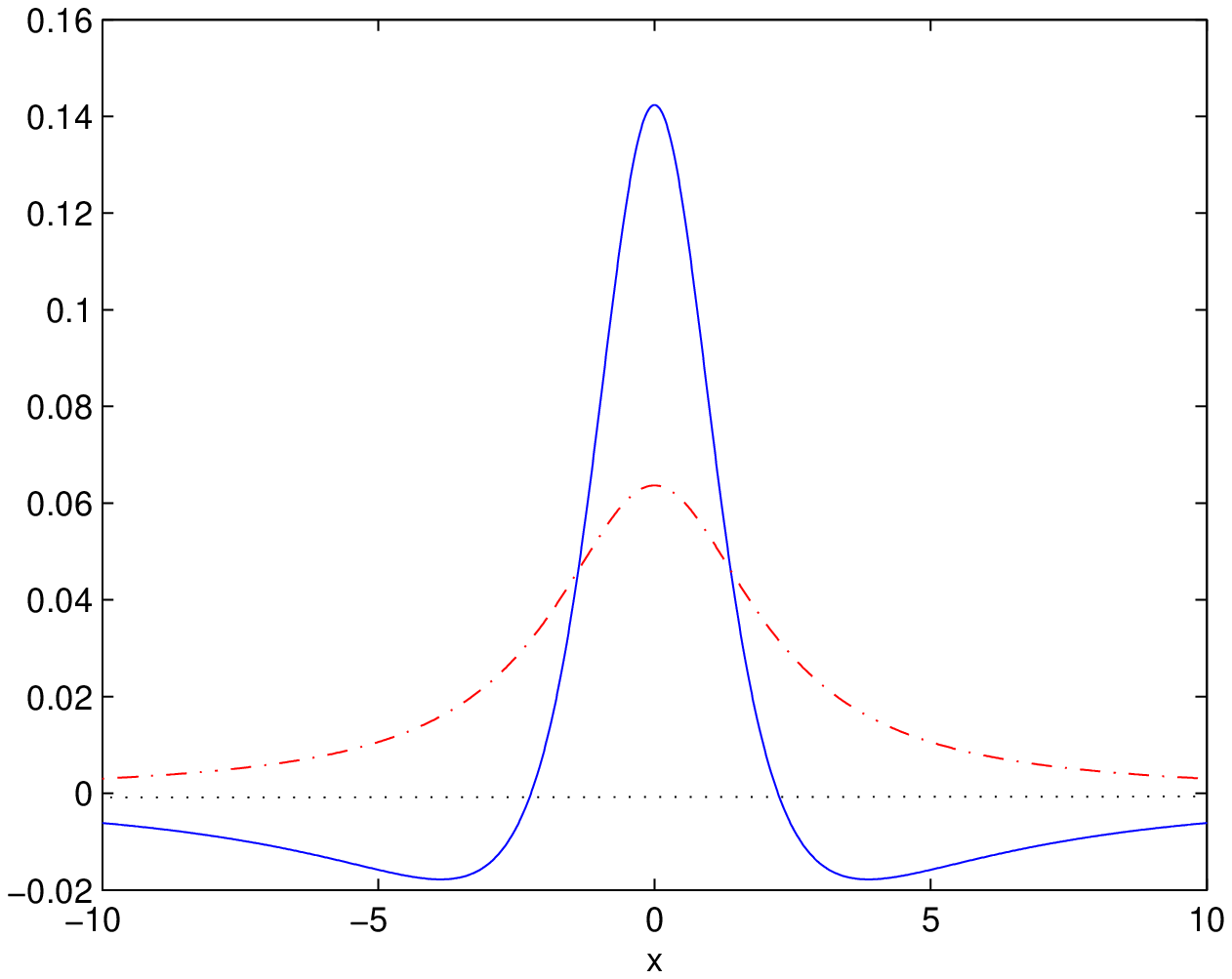}\vspace{0.2cm}
\caption{The corresponding figure of expression (\ref{Liudecomp}).
The Cauchy law (first term) is on dashed line while solid line
represents the second term with negative part. The zero value is
drawn with dotted line} \label{liudecomp}
\end{figure}

By varying the peakedness parameter m, this difference is retained
systematically. Analytically, one can also decompose the modified
Liu model to study its property. As an example, taking m=6, the
Liu PDF writes as:
\begin{equation}
\phi_L^6(x)= \frac{1}{\pi} \frac{\sqrt{5}}{5+x^2} + \frac{1}{\pi}
\frac{\sqrt{5}(5-x^2)}{(5+x^2)^2} \label{Liudecomp}
\end{equation}
where the first term is a Cauchy law $S_{1}(0,\sqrt{5},0)$. As
shown from the Fig.\ref{liudecomp}, the second term possesses a
significant positive values at the middle and the medium parts of
the probability law. The magnitude of these positive values are
responsible of the height of Liu PDF and the departure from the
stable law at the medium range values. The negative values of the
second term of (\ref{Liudecomp}) explain the similarity of the
asymptotic behavior of both laws observed in Fig.\ref{liucompare}.
For higher values of m, the same expansion holds with a Cauchy law
$S_{1}(0,\sqrt{m-1},0)$ as a first term but the second term
becomes a sum of rational functions whose integral vanishes to
ensure the normalization of the PDF.

According to the probability theory, the modified Liu PDF belongs
to the family of the Generalized Student or T-distribution whose
expression writes as:
\begin{equation}
\phi_S^m(x)=
\dps\frac{\Gamma(m+\frac{1}{2})}{\Gamma(\frac{m}{2})\quad\Gamma(\frac{1}{2})}
\left(\frac{a}{(x^2 +a^2)^{m+\frac{1}{2}}}\right) \label{student}
\end{equation}
where m is now called freedom degrees and a is a scale parameter.
The characteristic function of the T-distribution is given by:
\begin{equation}
\psi_S^m(t)= \quad2\dps \frac{|\mbox{at}|^{\frac{m}{2}}\quad \dps
K_{\frac{m}{2}}(|\mbox{at}|)} {2^{\frac{m}{2}} \Gamma(\frac{m}{2})
} \label{charstudent}
\end{equation}
where $K_i$ is the modified Bessel function. From (\ref{f_c}) and
(\ref{charstudent}), one can evaluate the difference between the
stable and the T-distribution laws. As said in previous section,
the stable laws decay as $|x|^{-\alpha-1}$ with $\alpha<2$. Here,
it may be shown that the T-distribution laws have the same
asymptotic behavior as $|x|^{-m-1}$ for all m$>$0. Note that as a
consequence, the T-distribution laws are more accurate than the
stable laws to approximate the near-Gaussian situation. Another
significant difference concerns the variance of the laws: for all
m$>$2, the T-distribution laws have finite variance given by
$\sigma^2=\frac{a^2}{m-2}$ while the stable laws do not possess
finite variance except only for $\alpha=2$. From the practical
viewpoint, this lack of variance may be of crucial importance when
one has to compute statistical values as the mean square slope
that is necessary to evaluate the radar backscatter cross section.

Finally, from the expression (\ref{student}), it appears that the
T-distribution does not include a skewness property while the
stable laws are naturally able to express the skewness effect.
Recalling the importance of the skewness effect on the dynamic of
the surface waves, this point may have significant consequences on
the accuracy of the slope distribution model.

\section{Conclusion}
The main purpose of this study was to highlight the possible use
of $\alpha$-stable laws for water surface slopes representation.
As said in the introduction of this paper, the fundamental
argument on the use of $\alpha$-stable laws stems first from
experimental observations of the main property of the wind driven
water surface waves: a possible fractal representation, related to
a spectral power law, nonlinear characters of the wind wavefield
resulting in an asymmetry and a modulational instability of the
surface, and a randomness of the slope, such as these abrupt
changes shown here in section 3.

Considering these strongly disordered behaviors that may be found
on slope variations, $\alpha$-stable laws seem \`a priori to be
potential candidates for this objective. Starting from this idea,
the study propose to apply the stable laws to represent the
probability distribution of the slopes of wind driven water
surface from laboratory experiments. It is found that the stable
laws are well adapted to model the random characteristic of the
slopes data in different cases of experiment configurations. From
the proposed cases, the four stable parameters behave in good
agreement with the effects of different physical processes
observed and expected in the experiments. In particular, the
stability index traduces fairly the non-Gaussianity of the slopes
in the same sense as the Stokes and/or the modulational
instability effects. Also the skewness parameter reproduces well
the asymmetry tendency of the slopes under various situations.
However, at extreme experiment condition (large fetch and high
wind speed), the stable laws fail to reproduce the behavior of the
upper part of the slopes.

From more theoretical consideration, comparisons with the
Gram-Charlier and the Liu et al\cite{Liu}  models are carried out.
It appears that the stable law differs completely from the
Gram-Charlier model as well as in its mathematical structure than
from the characteristics of its results. Indeed, one can recall
that the Gram-Charlier model at its usual truncation order (third
or fourth order) is part of the near-Gaussian approximation
models, while the stable law is totally non-Gaussian except only
when the stability index reaches its maximum value.

The comparison with the Liu model shows more similarities.
Recalling that Liu model belongs to the T-distribution family, it
is well established that both models are part of the so-called
"heavy tail probability laws". They are both infinitely divisible
probability laws but the T-distribution is not stable. This means
that the sum of independent identically T-distributed variables
does not follow the same law unlike the stable law that it is
always the case.

From mathematical viewpoint such stability property lies onto
fundamental characteristic based on the generalization of the
central limit theorem (see Samorodnitsky and Taqqu\cite{Samor})
and traduces the fact that the $\alpha$-stable laws constitute
limit laws (or attractor laws) for all additive random processes.
As a special case, we recall that for $\alpha=2$, from the central
limit theorem, the Gaussian law is a limit law for all additive
finite variance random processes. This is the profound
justification of the popularity of the Gaussian model in many
physical domains. In the realm of water surface waves, the random
wavefield is usually considered as the sum of infinite number of
random components. Owing to the weakness of the interactions
between components, the corresponding motion of each component may
be regarded as independent (see for example Phillips \cite{Phil}).
Consequently, under the generalisation of the central limit
theorem, the choice of stable law is then theoretically justified
with a clear mathematical basis.

With regard to the practical viewpoint, the differences between
the two laws are more important and have to be stressed here with
respect to the study objective. These differences lie on four
points concerning the closed form of the PDF, the finite variance
value, the skewness representation and the interpretation of the
physical processes through the PDF parameters.

As said in previous section, the T-distribution has a closed form
of the PDF for all values of its parameters .
There are only three cases of stable distributions for which a
closed form expression of the PDF is known. This lack of closed
form is apparently an inconvenient property of the stable law.
However, as shown in this study, there exists robust techniques to
implement the integral form of the stable PDF (see Zolotarev
\cite{zol}, Weron\cite{Weron}, Nolan\cite{Nolan}). Other expansion
techniques based on the Bergstr\"{o}m series are also available to
evaluate the stable PDF.

For physical applications, an appeal to statistical quantities
associated with the PDF model would be useful. For example, for
the sea surface remote sensing problem, it is known that the mean
and the variance of the slopes are of importance. This is
historically related to the assumption of near-Gaussian state in
the development. Recalling that T-distribution has finite variance
for all freedom degrees greater than 2. T-distribution appears as
a well adapted choice for this purpose. However, the
T-distribution and the stable distribution constitute fully
non-Gaussian alternative and then fail to belong to the category
of quasi-Gaussian solutions. In our opinion, an extension of the
remote sensing or the hydrodynamical relationships in the sense as
developed by Gu\'erin\cite{guerin} is required for a fully
non-Gaussian solution.

As far as non-Gaussian random water surface is concerned, the
skewness is undoubtedly one of its most important properties. The
stable distribution has asymmetric form that allows one to include
naturally the skewness. This is not the case of the T-distribution
for which no extension to asymmetric case is previewed. Empirical
asymmetric formulation has to be found (see Liu et al\cite{Liu}).

Before ending the paper, we focus on the interpretation of the
physical processes through the probability distribution model. As
we have shown in this study, the four stable parameters are able
to traduce correctly the effects of physical processes as the
asymmetry and the kurtosis of the slopes or also its level of
non-Gaussianity. Use of $\alpha$-stable distribution instead of
the traditional Gaussian or quasi-Gaussian distributions, combined
to hydrodynamical laws will be of great help on understanding the
physical behavior of the water wave slopes. In this sense,
extension of the present work to the space representation of water
wave slopes by the use of multivariate $\alpha$-stable
distribution has to be carried out and constitutes our prospective
task. An other way to be explored is also the use of truncated
stable law that allows one to obtain finite values of statistical
moments.

{\bf Acknowledgements} The authors are particularly grateful to
Yuri Stepanyants and to Fred Ramamonjiarisoa for their helpful
discussions on the first draft of this manuscript. The authors
thank also E. Michel for his help to improve the text.

\end{document}